\documentclass[floatfix,twocolumn,showpacs,amsmath,amssymb]{revtex4}
\usepackage{graphicx}
\usepackage{times}
\usepackage{hyperref}

\newcommand{\fwidth}{.45\textwidth}  

\begin{document} 

\title{Newtonian vs. relativistic chaotic scattering}
\author{Juan~M.~Aguirregabiria, \email{juanmari.aguirregabiria@ehu.es}} 
\affiliation{Theoretical Physics, 
The University of the Basque Country, \\
P.~O.~Box 644,
48080 Bilbao, Spain}
\email{juanmari.aguirregabiria@ehu.es} 

\date{\today}

\begin{abstract}
It is shown that Newtonian mechanics is not appropriate to compute approximate individual
trajectories with small velocities in chaotic scattering. However, some global
properties of the dynamical system, such as the dimension of the 
non-attracting chaotic invariant set, are more robust and the Newtonian approximation
provides reasonably accurate results for them in slow chaotic scattering.
\end{abstract}




\maketitle

\section{\label{sec:intro}Introduction}

It has been recently pointed out that, when the evolution is chaotic, the trajectories predicted by Newtonian mechanics 
are not good approximations to the special-relativistic trajectories, even for very small velocities,
\cite{lan1,lan2,lan3}. The problem considered
in Refs.~\cite{lan1,lan2,lan3} is the evolution of periodically delta-kicked systems
with or without damping. The purpose of this letter is twofold:
we want to extend the previous conclusion to another area of deterministic chaos 
(chaotic scattering) but also to point out that, for slow motion, 
Newtonian mechanics can provide
good approximations for global characteristics of chaotic
systems, such as the structure and dimension of the 
non-attracting chaotic invariant set.

We will consider a relativistic particle moving in an external potential energy $V(\mathbf{r})$.
If particle's mass, position and linear momentum are $m$, $\mathbf r$ and $\mathbf p$, respectively,
 the Hamiltonian (i.e., the total energy) is $H=\sqrt{m^2c^4+c^2\mathbf{p}^2}+V(\mathbf{r})$
 and Hamilton's canonical equations read as follows:
 \begin{equation}\label{eq:hamilton}
\dot{\mathbf{r}}=\frac{\mathbf{p}}{m\gamma},\quad \dot{\mathbf{p}}=-\nabla V(\mathbf{r}),  
 \end{equation}
where the Lorentz factor is given by
\begin{equation}\label{eq:lorentz}
\gamma\equiv\sqrt{1+\frac{\mathbf{p}^2}{m^2c^2}}=\left(1-\frac{\dot{\mathbf{r}}^2}{c^2}\right)^{-1/2}.
\end{equation}
The Newtonian equations of motion are recovered by putting $\gamma=1$, so that one
might think they are a good approximation for slow motion, where $\gamma\approx1$.
We will discuss in the next section to what extent is right this assumption
in the case of chaotic scattering.

\section{\label{sec:scat}Chaotic scattering}

Let us consider a particle moving in the plane $(x,y)$ under
the force derived from the potential \cite{bleher,Ott}
\begin{equation}
V(x,y)=x^2y^2\,e^{-\left(x^2+y^2\right)}. 
\end{equation}
The particle is sent from $(x,y)=(-\infty,b)$ with initial
velocity $(\dot x,\dot y)=(0,v)$ and after some time emerges from the
 scattering zone moving freely along
a direction forming an angle $\Phi$ with the positive $x$ axis. 
The initial velocity $v$ is computed from the conserved energy $E=H-mc^2$.
We choose a system of units in which $m=c=1$
and a value of the energy $E=0.05V_0$ ---where $V_0=e^{-2}$
is the height of the four maxima of the potential located at $(x,y)=(\pm1,\pm1)$---,
so that the motion is slow, since the Lorentz factor $\gamma$ is always smaller than $1.007$.

\begin{figure}
\includegraphics[width=\fwidth]{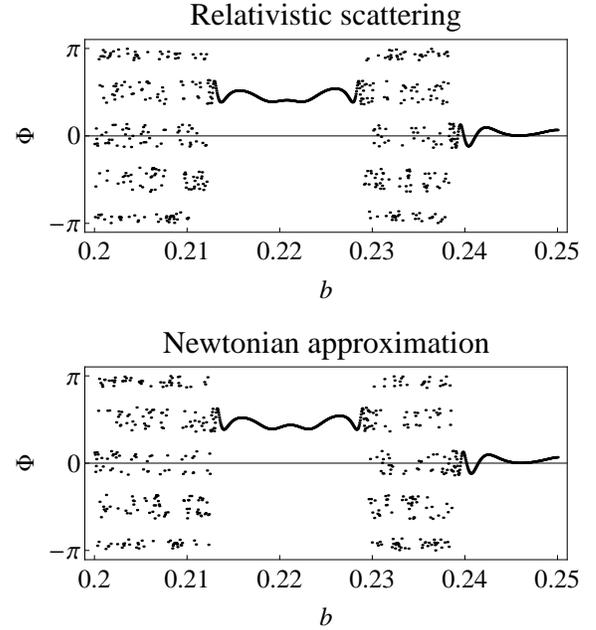}
\caption{Scattering angle $\Phi(b)$ for the relativistic and Newtonian equations of motion. \label{fig1}}
\end{figure}

In Fig.~\ref{fig1} we have depicted 1000 values of the scattering function $\Phi(b)$,
in the range $[0.2,0.25]$ of the impact parameter $b$, obtained by numerical integration \cite{DS}
of the relativistic equations of motion (\ref{eq:hamilton}) and their Newtonian
approximation.

First of all, one can see that while Newtonian mechanics gives reasonably approximate 
results when the scattering function is smooth, when the latter is discontinuous
the predictions of relativistic and Newtonian equations of motions are very often
completely different. This extends to chaotic scattering the
claim in Refs.~\cite{lan1,lan2,lan3} that Newtonian equation of motions are not a good
approximation when \emph{individual} chaotic trajectories have to be computed.

\begin{figure}[ht]
\includegraphics[width=\fwidth]{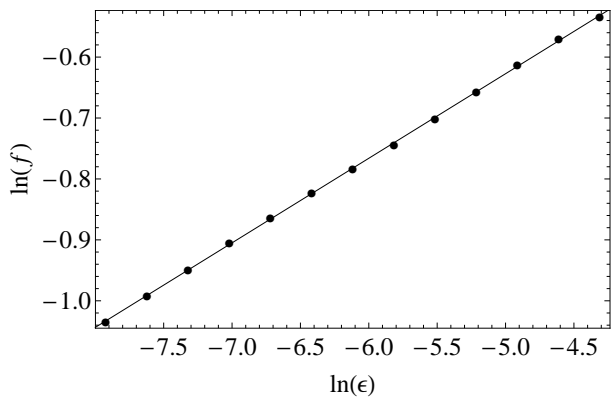}
\caption{Log-log plot of some points of the uncertainty function $f(\epsilon)$
and a linear fit. \label{fig2}}
\end{figure}

However, although the positions of individual points in the regions
of chaotic scattering may be very different, the position of the discontinuity
points along the $b$ axis seems to be very similar in both cases. This statement can be made quantitative if
one computes the dimension of the Cantor set of discontinuity points by using the
method of the uncertainty exponent \cite{bleher,Ott}. We have computed the
scattering function for $2^{22}$ values of $b\in[0.2,0.25]$ and then selected 
a small value $\epsilon$. We say that a value $b$ is \emph{certain} if
for the three impact parameters $b-\epsilon$, $b$ and $b+\epsilon$ the scattering
is upward or if it is downward for the three values. We repeat this for many values of $b$ and
call $f(\epsilon)$ to the proportion of  uncertain values of $b$. Figure~\ref{fig2}
show a log-log plot of several values of $f(\epsilon)$ along with a good linear fit, so that
$f(\epsilon)\sim e^\alpha$ with a uncertainty exponent given by the slope of the straight line: 
$\alpha\approx 0.139$. In consequence \cite{bleher,Ott}, 
the fractal dimension of the set of discontinuity points is 
$D_0=1-\alpha\approx 0.861$. The corresponding plot for the
Newtonian approximation looks nearly identical with $\alpha\approx 0.141$. This 
shows that the dimension of the chaotic invariant set is well approximated
when computed with Newtonian mechanics, if the velocities are small enough.  

\section{\label{sec:comments}Final comments}

In the previous section we have used Newtonian mechanics in a consistent way, so that for each
impact parameter $b$ the initial velocity $v$ is computed using the Newtonian definition for
the kinetic energy. Since the latter is different in special relativity, slightly different
initial velocities were used in the Newtonian and the relativistic simulations. If one 
uses the same initial velocities in both cases (which implies slightly 
different energies), the results are very similar to those discussed above. For instance,
one gets $\alpha\approx 0.142$.

We have shown that also in chaotic scattering Newtonian mechanics fails to be a good approximation
to compute individual trajectories. Even for slow motion, the little relativistic corrections
to the equations of motion, which remain small in non-chaotic scattering, 
are greatly amplified by chaotic scattering during the long time
in which the trajectory remains close to the non-attracting invariant set. This extends to chaotic scattering the
conclusions reached in Refs.~\cite{lan1,lan2,lan3}.
However, we have also shown that some global properties, such as the
structure and dimension of the set of discontinuity points (and, thus, the dimension
of the non-attracting chaotic invariant set) are more robust and Newtonian mechanics
provides reasonably accurate approximations to them if the motion is slow enough.

\section*{Acknowledgments}
This work was supported by the Basque Government (Department of Education, Universities and Research)
(Research Grant~GIC07/51-IT-221-07).

\section*{References}

\end{document}